\documentclass[aps,onecolumn,prd,amssymb,eqsecnum,nofootinbib,superscriptaddress,showpacs,floatfix,amsmath]{revtex4}
\newlength{\picwidth}
\usepackage[dvips]{graphicx}
\usepackage{dcolumn}
\usepackage{amsmath}
\usepackage{amssymb}
\usepackage{amsfonts}
\usepackage{bm}

\def\be{\begin{equation}}
\def\ee{\end{equation}}
\def\ba{\begin{eqnarray}}
\def\ea{\end{eqnarray}}
\def\href{H_{0}}

\def\rfrw{r_{\rm FRW}}

\def\dtzz{{dt_0\over dz}}

\def\zc{z_{\rm crit}}
\def\OF{\Omega_{\rm FRW}^{1/2}}
\def\vinfty{V_\infty}
\def\rfinfty{r_{{\rm FRW},\infty}}
\def\kfrw{K_{\rm FRW}}
\def\rtil{{\tilde r}}

\def\heff{H_{\rm eff}}
\def\weff{w_{\rm eff}}
\def\pp{n}
\def\Rp{\partial R/\partial r}
\def\Rdotp{\dot R'}
\begin{document}
\title{Mimicking dark energy with Lema\^{i}tre-Tolman-Bondi models: weak central
singularities and critical points}

\author{R. Ali Vanderveld}
\affiliation{Center for Radiophysics and Space Research, Cornell University, Ithaca, NY 14853}
\author{\'{E}anna \'{E}. Flanagan}
\affiliation{Center for Radiophysics and Space Research, Cornell University, Ithaca, NY 14853}
\affiliation{Laboratory for Elementary Particle Physics, Cornell University, Ithaca, NY 14853}
\author{Ira Wasserman}
\affiliation{Center for Radiophysics and Space Research, Cornell University, Ithaca, NY 14853}
\affiliation{Laboratory for Elementary Particle Physics, Cornell University, Ithaca, NY 14853}
\date{\today}
\begin{abstract}

There has been much debate over whether or not one could explain the
observed acceleration of the Universe with inhomogeneous cosmological
models, such as the spherically-symmetric Lema\^{i}tre-Tolman-Bondi
(LTB) models. It has been claimed that the central observer in these models
can observe a local acceleration, which would contradict
general theorems.  We resolve the contradiction by noting
that many of the models that have been explored contain a weak
singularity at the location of the observer which makes them unphysical.  In the absence of this
singularity, we show that LTB models must have a positive central
deceleration parameter $q_{0}$, in agreement with the general
theorems.  We also show that it is possible to achieve a negative apparent deceleration parameter at nonzero redshifts in LTB models that do not contain this singularity.
However, we find other singularities that tend to arise in LTB models
when attempting to match luminosity distance
data, and these generally limit the range of redshifts for which
these models can mimic observations of an accelerating Universe. Exceptional models do exist that can extend to arbitrarily large redshift without encountering these pathologies, and we show how these may be constructed. These special models exhibit regions with negative effective equation of state parameter, which may fall below negative one, but we have failed to find any singularity-free models that agree with observations. Moreover, models based on dust-filled LTB metrics probably fail to reproduce observed properties of large scale structure.

\end{abstract}
\maketitle

\section{Introduction}

The Universe appears to be expanding at an accelerating rate.  This has been deduced from luminosity distance measurements of Type Ia supernovae, which appear dimmer than one would expect based on general relativity without a cosmological constant \cite{Riess, Perlmutter}, and from measurements of the current matter density $\Omega_{M}\approx 0.27$, which is too small to close the Universe as required by cosmic microwave background radiation (CMBR) observations \cite{Bennett}.  Many explanations for this discrepancy have been put forward, most entailing a modification of general relativity on cosmological scales or the addition of a new field with exotic properties, called ``dark energy"; for reviews, see \cite{Peebles, Carroll}.

However, there have also been attempts to explain this seemingly anomalous cosmic acceleration as a consequence of inhomogeneity rather than modified gravity or dark energy.  Many recent papers have claimed that superhorizon density perturbations, left over from inflation, could backreact and drive the acceleration of our local universe \cite{Kolb1, Barausse}.  Problems with this argument have been pointed out \cite{Flanagan, Geshnizjani, Hirata, Rasanen1}, and it does not appear to be a plausible alternative.   

In this paper we will only consider subhorizon perturbations, which have also been considered as a possible way to obtain an accelerating universe.  Kolb {\it et al.} use second order perturbation theory \cite{Kolb2, Kolb3} to suggest that density perturbations could backreact to cause accelerated expansion without the need to introduce any form of dark energy, which is an appealing prospect. R\"{a}s\"{a}nen \cite{Rasanen2} and Notari \cite{Notari} had earlier explored this claim by looking at average expansion parameters in order to argue that we could measure acceleration due to this effect. In opposition to this, Siegel and Fry \cite{Siegel} and Ishibashi and Wald \cite{Ishibashi} have argued that our universe is very accurately described with a Newtonianly perturbed Friedmann-Robertson-Walker (FRW) metric, and, treated as such, it does not permit accelerated expansion due to perturbations.

Since it has proven to be quite complicated to analyze the full three-dimensional backreaction problem analyzed by Kolb {\it et al.}, a useful class of models to explore are the spherically-symmetric, yet inhomogeneous, Lema\^{i}tre-Tolman-Bondi (LTB) \cite{Bondi} cosmological models, containing only cold dark matter, or ``dust", and wherein it is often, but not always, assumed that we live at the symmetry center. In this way, we can confront the simpler and more general question: Are there any models based on general relativity and cold dark matter which can match the observations? We cannot completely adress this question with LTB models, which are unrealistic since they place us near the center of the Universe,
but these models are nevertheless useful toy models to address this general
question. More specifically, in the LTB models we can ask if a centrally
located observer can mistakenly interpret astronomical observations of redshifts and luminosity distances
as requiring acceleration of the expansion of the Universe. We find that the answer is ``yes", and this implies that the mechanism studied by Kolb {\it et al.} is somewhat more plausible and requires more study. 

Other papers have used LTB models in analyzing whether or not subhorizon perturbations could backreact and drive accelerated expansion. Nambu {\it et al.} take averages to find effective expansion parameters of specific illustrative example models \cite{Nambu}, Moffat looks at examples \cite{Moffat}, Mansouri constructs a model that consists of a local LTB patch which is embedded into a background FRW spacetime \cite{Mansouri}, and Chuang {\it et al.} numerically produce examples of LTB models with apparent acceleration \cite{CGH}.  Alnes {\it et al.} \cite{Alnes1} argue against acceleration, but only by looking at a class of example models. 

It has been claimed that it is possible to find LTB cosmological models that have $q_{0}<0$, where $q_{0}\equiv q(z=0)$ is the deceleration parameter measured by the central observer.  However, there are general theorems that prohibit such behavior \cite{Flanagan, Hirata}.  In Section II, we will first give a general review of LTB models and then we will discuss this contradiction and its resolution: there is a local singularity at the symmetry center of models with $q_{0}<0$, corresponding to a non-vanishing radial central density gradient and divergent second derivatives of the density. We will prove that excluding this singularity will necessarily lead to a positive value for $q_{0}$. This singularity is not taken into account in any of the above papers, and most of them look at models which are singular at the center \cite{Nambu, Moffat, Mansouri, CGH}. 

We will also show that it is possible to construct models without a central singularity in which one would measure negative deceleration parameters $q(z)$, and therefore would measure regions of acceleration, at nonzero redshifts $z$. We will do this by choosing the LTB model and computing the resulting luminosity distance and ultimately $q(z)$; we call this the ``forward problem". As we discuss in Section II below, LTB models are characterized by two free functions of radius $r$, a bang time function $t_0(r)$ and an energy function $E(r)$. We focus on LTB models with zero energy functions but non-zero bang time functions, because we do not expect the former to produce acceleration.  This is because, as we will show, the energy function is associated with the growing mode of linear theory, whereas the bang time function is associated with the shrinking mode.

In Section III, we will explore the ``inverse problem", where one chooses the luminosity distance as a function of redshift and then attempts to find a corresponding LTB model, which may or may not exist. Here, too, we only consider $E(r)=0$. We show that there are numerous pitfalls to this method, as other singular behaviors arise which generally limit the range of redshifts for which this class of models could reproduce the observed supernova data. For a given luminosity distance $D_{L}(z)\equiv r_{FRW}(z)(1+z)$, there is a critical redshift $\zc$ where $d\ln r_{FRW}(z)/d\ln(1+z)=1$.  For almost all choices of $D_{L}(z)$, any attempt to find a corresponding zero energy LTB model will fail at some redshift smaller than $\zc$ when a singularity is encountered. There are exceptions which pass through a ``critical point" at $z=\zc$, the simple FRW model being one obvious example of such a ``transcritical" solution. We show how others may be constructed. These models show redshift domains with enhanced deceleration as well as acceleration, but do not appear to be consistent with observational data on $D_{L}(z)$.  

Several papers have already computed how the dependence of the luminosity distance on redshift is distorted in LTB models due to purely radial inhomogeneities, and have claimed that we could be tricked into thinking that we are in a homogenous accelerating universe when we are really in a dust-dominated inhomogeneous universe \cite{Mustapha, Sugiura, Celerier, INN}. However, this claim has not until now been correctly justified, since all previous papers neglected
the central singularity and the critical point.

\section{The Forward Problem}

\subsection{Lema\^{i}tre-Tolman-Bondi Models}

Using the notation of C\'{e}l\'{e}rier \cite{Celerier}, the LTB spacetime \cite{Bondi} has the line element
\be
ds^{2}=-dt^{2}+\frac{R'^{2}(r,t)}{1+2E(r)}dr^{2}+R^{2}(r,t)\left(d\theta^{2}+\sin^{2}\theta d\phi^{2}\right)
\ee
where primes denote derivatives with respect to the radial coordinate $r$, and $E(r)$ is a free function, called the ``energy function". 
We define the function $k(r)$ by $k(r)\equiv-2E(r)/r^2$. If $k(r)=0$, the Einstein equations admit the solution
\be
R(r,t)=\left(6\pi G\tilde{\rho}\right)^{1/3}r\left[t-t_{0}(r)\right]^{2/3}~,
\label{Rflat}
\ee
where $t_{0}(r)$ is another free function, often referred to as the ``bang time" function, and $\tilde{\rho}$ is a fixed parameter.  If $k(r)<0$ for all $r$, we have the parametric solution
\ba
R&=&\frac{4\pi G\tilde{\rho}r}{-3k(r)}\left(\cosh u-1\right) \nonumber\\
t-t_{0}(r)&=&\frac{4\pi G\tilde{\rho}}{3\left[-k(r)\right]^{3/2}}\left(\sinh u-u\right)~,
\label{para1}
\ea
and if $k(r)>0$ for all $r$, we have the solution
\ba
R&=&\frac{4\pi G\tilde{\rho}r}{3k(r)}\left(1-\cos u\right) \nonumber\\
t-t_{0}(r)&=&\frac{4\pi G\tilde{\rho}}{3\left[k(r)\right]^{3/2}}\left(u-\sin u\right)~.
\label{para2}
\ea 
These are Eqs.\ (18), (19), and (20) of C\'{e}l\'{e}rier \cite{Celerier}, but specialized to the choice $M(r) = 4 \pi r^3 \tilde{\rho}/3$ by choosing the radius coordinate appropriately, where $M(r)$ is the mass
function used by C\'{e}l\'{e}rier, and where $\tilde{\rho}$
is a constant. \footnote{Note that the mass function $M(r)$ which
appears in Bondi \cite{Bondi}, which we denote by $M_B(r)$, is related to 
C\'{e}l\'{e}rier's $M(r)$ by $M'_B(r) = M'(r)/\sqrt{1 + 2 E(r)}$, and so our radial coordinate specialization 
in Bondi's notation is $M_B'(r) = 4 \pi r^2 \tilde{\rho} / \sqrt{1 -2 k(r) r^2}$.}
The energy density of the matter in these models is given by
\be
\rho(r,t)=\frac{\tilde{\rho}r^{2}}{R'R^{2}}~.
\label{density}
\ee
We define $\rho_0(t) = \rho(0,t)$ to be the central density, and from Eqs.\ (\ref{Rflat})-(\ref{para2}) we find
\be
\rho_0(t)=\frac{1}{6\pi G\left[t-t_0(0)\right]^2}~.
\label{triplestar}
\ee
Throughout this paper we will restrict attention to an observer located at $r=0$ and at $t=t_o$,
where $t_o$ is the observation time, not to be confused with the bang time $t_0(r)$. We also choose units such that $\tilde{\rho} = \rho_0(t_o)$, and we choose the origin of
time such that $t_0(0)=0$.
A light ray directed radially inward follows the null geodesic
\be
dt=-\frac{R'(r,t)}{\sqrt{1-k(r)r^{2}}}dr
\label{null}
\ee
and has a redshift given by
\be
\frac{dz}{dr}=\left(1+z\right)\frac{\dot{R}'\left[r,t(r)\right]}{\sqrt{1-k(r)r^{2}}}
\label{redshift}
\ee
where overdots denote partial derivatives with respect to time and where $t(r)$ is evaluated along light rays that are moving radially inward according to Eq.\ (\ref{null}).  Equations (\ref{density}) and (\ref{redshift}) give us two important restrictions on the derivatives of $R(r,t)$: (i) in order for the density to remain finite, we require $R'>0$, which excludes shell-crossing, and (ii) in order to have a monotonically increasing $z(r)$, we require $\dot{R}'>0$.

The luminosity distance measured by the observer at $r=0$ and at $t=t_o$ is given by \cite{Celerier}
\be
D_{L}(z)=\left(1+z\right)^{2}R~,
\label{lum}
\ee
where $z$ and $R$ are evaluated along the radially-inward moving light ray. It is not obvious how to define the deceleration parameter in an inhomogeneous cosmology, and Hirata and Seljak \cite{Hirata} explore several definitions.  In this paper, we restrict our attention to the deceleration parameter that would be obtained from measurements of luminosity distances and redshifts assuming a spatially flat FRW cosmology. \footnote{More generally, an observer might fit data on $D_{L}(z)$ to FRW models with arbitrary spatial curvature, including flat ones.} We can deduce the effective Hubble expansion rate $H(z)$ of the flat FRW model which would yield the same luminosity distances by inverting the FRW relation
\be
D_{L}(z)=\left(1+z\right)\int^{z}_{0}\frac{dz'}{H\left(z'\right)}
\ee
to find
\be
H(z)=\left[\frac{d}{dz}\left(\frac{D_{L}(z)}{1+z}\right)\right]^{-1}~.
\label{hubble}
\ee
We can then calculate the associated deceleration parameter
\be
q(z)=-1+\left[\frac{1+z}{H(z)}\right]\frac{dH(z)}{dz}
\label{decel}
\ee
and the effective equation of state parameter 
\be
\weff(z)\equiv\frac{2}{3}\left[q(z)-\frac{1}{2}\right]=\frac{2(1+z)}{3}\frac{d}{dz}\ln\left[\frac{H(z)}{(1+z)^{3/2}}\right]~.
\ee
If we know $t_{0}(r)$ and $E(r)$, then we can find $R(r,t)$ very simply by using the appropriate solution above, chosen from Eqs.\ (\ref{Rflat}), (\ref{para1}), and (\ref{para2}). We then solve the differential equations (\ref{null}) and (\ref{redshift}) to find $t(z)$ and $r(z)$, starting from the initial conditions $r=0$ and $t=t_o$. We insert these $t(z)$ and $r(z)$ into the right hand side of Eq.\ (\ref{lum}) to obtain $D_{L}(z)$, and then use Eqs.\ (\ref{hubble}) and (\ref{decel}) to find $H(z)$ and $q(z)$. We will use this procedure later in this section with a class of models as an illustrative example.

\subsection{The Weak Singularity at $r=0$}

There have been many claims that there exist LTB cosmological models in which $q_{0}\equiv q(z=0)<0$ \cite{Nambu, Moffat, Mansouri, CGH, Mustapha, Sugiura, Celerier, INN}. For example, Iguchi {\it et al.} \cite{INN} look at two different classes of LTB models: (i) models with $k(r)=0$ and a pure ``BigBang time inhomogeneity" and (ii) models with $t_{0}(r)=0$ and a pure ``curvature inhomogeneity".  In either case, they try to reproduce the luminosity distance function of a flat FRW universe with a matter density $\Omega_{M}=0.3$ and a cosmological constant density $\Omega_{\Lambda}=0.7$, namely
\be
D_{L}(z)=\frac{1+z}{H_{0}}\int^{z}_{0}\frac{dz'}{\sqrt{\Omega_{M}\left(1+z'\right)^{3}+\Omega_{\Lambda}}}~.
\label{FRW}
\ee
They appear to be successful up until they find $R'<0$ or $\dot{R}'<0$ at a redshift $z\sim 1$ (we will discuss these pathologies in the next section). Thus, they appear to successfully find models where $q_{0}<0$.

On the other hand, the local expansions of Flanagan \cite{Flanagan} and of Hirata and Seljak \cite{Hirata} show that $q_{0}$ is constrained to be positive for arbitrary inhomogeneous dust-dominated cosmologies that are not necessarily spherically-symmetric.  In particular, Flanagan expands the luminosity distance as
\be
D_{L}=A(\theta,\phi)z+B(\theta,\phi)z^{2}+{\cal O}(z^{3})~,
\ee
where $\theta$ and $\phi$ are spherical polar coordinates as measured in the local Lorentz frame of the observer. He then defines the central deceleration parameter as
\be
q_{0}\equiv 1-2H_{0}^{-2}\langle A^{-3}B\rangle~,
\ee
where angle brackets denote averages over $\theta$ and $\phi$, and $H_0=\langle A^{-1}\rangle$.  Using local Taylor series expansions and assuming that the pressure is zero, he finds
\be
q_{0}=\frac{4\pi}{3H_{0}^{2}}\rho+\frac{1}{3H_{0}^{2}}\left[\frac{7}{5}\sigma_{\alpha\beta}\sigma^{\alpha\beta}-\omega_{\alpha\beta}\omega^{\alpha\beta}\right]
\ee
where $\sigma_{\alpha\beta}$ and $\omega_{\alpha\beta}$ are the shear and vorticity tensors. The first term of this expression is obviously positive, and the terms in the brackets vanish in LTB models by spherical symmetry.  Thus there is a contradiction: general theorems prove that $q_{0}$ is positive in these inhomogeneous models, whereas the analysis of specific examples appears to show that it is possible to construct models in which $q_{0}$ can be negative.  Here we present the resolution of this contradiction, that there exists a weak local singularity which is excluded at the start from the computations of Flanagan and Hirata and Seljak, but which is present in models giving $q_{0}<0$. We will show that the exclusion of this singularity inevitably leads to models with a positive $q_{0}$.

We expand the density (\ref{density}) to second order in $r$ as
\be
\rho(r,t)=\rho_{0}(t)+\rho_{1}(t)r+\rho_{2}(t)r^{2}+{\cal O}\left(r^{3}\right)~.
\label{rhoexp}
\ee
The weak singularity occurs when $\rho_{1}(t)$ is nonzero, in which case the gravitational field is singular since $\Box {\cal R}\to\infty$ as $r\to 0$, where ${\cal R}$ is the Ricci scalar. In other words, second derivatives of the density diverge at the origin, independent of where observers may be located.
This is true both in flat spacetime and in the curved LTB metric when we have a density profile of the form
(\ref{rhoexp}). The singularity is weak according to the classification scheme of the literature on general relativity \cite{Tipler}. This singularity is excluded from the start in the analyses of Flanagan \cite{Flanagan} and Hirata and Seljak \cite{Hirata} which assume that the metric is smooth.

We now determine the conditions for a weak singularity to occur. We define the variable
\be
a(r,t)=\frac{R(r,t)}{r}~;
\label{plus}
\ee
this is analogous to the FRW scale factor $a(t)$, in the sense the metric takes the form
\be
ds^{2}=-dt^{2}+a^{2}(r,t)\left\{\frac{\left[1+ra'(r,t)/a(r,t)\right]^{2}}{1-k(r)r^{2}}dr^{2}+r^{2}\left(d\theta^{2}+\sin^{2}\theta d\phi^{2}\right)\right\}~.
\ee
We expand this function as
\be
a(r,t)= a_{0}(t)+a_{1}(t)r+a_{2}(t)r^{2}+{\cal O}\left(r^{3}\right)~.
\label{artexp}
\ee
Comparing this to the formula (\ref{Rflat}) for $R(r,t)$, we find for the zeroth order expansion coefficient
\be
a_0(t)=\left[6\pi G\rho_0\left(t_o\right)\right]^{1/3}t^{2/3}~.
\label{doublestar}
\ee
We define $H_{0}=\dot{a}_0(t_o)/a_0(t_o)$, and our choice of units above imply $a_0(t_o)=1$. Using Eqs.\ (\ref{plus}) and (\ref{artexp}) in the expression (\ref{density}) for the density gives
\be
\rho(r,t)=\frac{\rho_0\left(t_o\right)}{a_0^2(t)}-4\frac{\rho_0\left(t_o\right)a_1(t)}{a_0^3(t)}r+{\cal O}\left(r^2\right)~.
\ee
Since $a_0(t)\neq 0$ by Eq.\ (\ref{doublestar}), we see that having a non-singular model requires $a_1(t)=0$, or equivalently $R''(r=0,t)=0$.

It is straightforward to see that if $a_{1}=0$, then $q_{0}\geq 0$, and that if $a_{1}(t)\neq 0$, then $q_{0}$ may be positive or negative. Note that the 
observer's measurement of $q_0$ from the data does not depend
on the observer's prior assumptions about spatial curvature, and so the following analysis of $q_0$ is sufficiently general and applies for arbitrary $k(r)$. If $a_{1}(t)=0$, then the angular size distance is $R(r,t)=ra_0(t)+
r^3a_2(t)+{\cal O}(r^4)$, where $r$ and $t$ are evaluated along the path followed by a radially directed light ray. Evaluating the
redshift for such a ray gives to lowest order $z=H_0r+{\cal O}(r^2)$. Thus, the angular size distance
is unaffected by density gradients up to terms of order $z^3$.
In other words, the standard expansion of the angular size distance $R\equiv D_{A}$ to 
order $z^2$,
\be
H_0D_A(z)=z-{1\over 2}z^2(3+q_0)+{\cal O}(z^{3})~,
\label{asd}
\ee
is completely determined by the evolution of the uniformly dense
core region of the expanding spherically symmetric model, where
the density is $\rho_0(t)=\rho_0(t_o)/a_0^3(t)$ from Eqs.\ (\ref{triplestar}) and (\ref{doublestar}), which is the density of dust
expanding with scale factor $a_0(t)$. Therefore, the effective
values of $q_0$ for such a model must lie
in the same range as are found for exactly uniform, dust dominated FRW models:
$q_0\geq 0$.  

We can gain further physical insight into the behavior of LTB models near $r=0$ by expanding the field equations in $r$, assuming (see Eq.\ (\ref{artexp}))
\be
a(r,t)=a_{0}(t)+a_{n}(t)r^{n}+\ldots~, 
\label{aexp}
\ee
and correspondingly
\be
k(r)=k_{0}+k_{n}r^{n}+\ldots~;
\label{kexp}
\ee
we show in Appendix \ref{proof2} that $a_1(t)=0$ corresponds to having $k_1=0$ via a direct analysis of the LTB solutions. Thus, for non-singular models, $n=2$ is the leading order correction to strict homogeneity near the center. The field equations are given in Bondi \cite{Bondi}, and in our notation
his Eq.\ (24) is
\be
{1\over 2}\left({\partial R(r,t)\over\partial t}\right)^2
-{4\pi G\rho_0\left(t_{o}\right) r^3\over 3R(r,t)}=-{1\over 2}k(r)r^2~.
\ee
Substituting $R(r,t)=ra(r,t)$ we find
\be
\left({\partial a(r,t)\over\partial t}\right)^2=
{8\pi G\rho_0\left(t_{o}\right)\over 3a(r,t)}-k(r)~.
\ee
Using the expansions (\ref{aexp}) and (\ref{kexp}) and equating like powers of $r$, we find
\ba
\dot a_0^2&=&{8\pi G\rho_0\left(t_{o}\right)\over 3a_0}-k_0
\equiv {H_0^2\Omega_0\over a_0}+H_0^2(1-\Omega_0) \nonumber\\
2\dot a_0\dot a_n&=&-{8\pi G\rho_0\left(t_{o}\right) a_n\over 3a_0^2}-k_n
=-{H_0^2\Omega_0 a_n\over a_0^2}-k_n~.
\label{lbteqns}
\ea
The first of Eqs.\ (\ref{lbteqns}) is exactly the same as the
Friedmann equation for the scale factor $a_0(t)$ in a universe with
arbitrary spatial curvature, subject to the single physical
requirement $\Omega_0\geq 0$. To solve the second equation,
notice that $\ddot a_0=-H_0^2\Omega_0/2a_0^2$, so rewrite it
as
\be
\dot a_0\dot a_n-\ddot a_0 a_n=\dot a_0^2{d\over dt}\left(
{a_n\over \dot a_0}\right)=-{k_n\over 2}~,
\ee
which has the solution
\be
a_n(t)=C\dot a_0-{k_n\dot a_0\over 2}\int_{0}^t{dt'\over \dot a_0^2(t')}~,
\label{atwosoln}
\ee
where $C$ is a constant.
Let us define $\delta_n(t)=a_n(t)/a_0(t)$; then Eq.\ (\ref{atwosoln}) becomes
\be
\delta_n(t)=CH(t)-{k_nH(t)\over 2}\int_{0}^t{dt'\over H^2(t')a_0^2(t')}
=CH(t)-{k_nH(t)\over 2}\int_{a_0(0)}^{a_0(t)}{da_0\over H^3(a_0)a_0^3}~,
\label{deltwosoln}
\ee
where 
\be
H\equiv{\dot a_0\over a_0}=H_0\sqrt{{\Omega_0\over a_0^3}+
{1-\Omega_0\over a_0^2}}~.
\ee
Comparing with results in Peebles \cite{Peebles2}, we see that the first term of Eq.\ (\ref{deltwosoln}) is just the
shrinking mode of linear theory, and the second is the growing mode.
The amplitude $C$ of the shrinking mode is related to the bang time function by
$t_0(r)=-Cr^n+\ldots$, and the growing mode amplitude $k_n$ is related to the lowest order energy perturbation $k_{n}r^{n}$. Note that this approximate solution holds for $a_{n}r^{n}<<a_{0}(t)$, i.e. for $0<r^{n}<<1/\delta_{n}$.

We have shown that for $n\geq 2$, the central value of $q_{0}$ is greater than or equal to zero.  We now compare with the mildly singular case with $n=1$. The evolutions of $a_{0}(t)$ and $a_{1}(t)$ are governed by Eqs.\ (\ref{lbteqns}) and (\ref{atwosoln}).
For this case, the low $z$
expansion of $D_A(z)$ depends on $a_1(t_o)$ and $k_1$, and the
effective value of $q_0$ near the origin becomes, from Eqs.\ (\ref{plus}), (\ref{asd}), and (\ref{aexp}),
\be
q_{0}=
{1\over 2}\Omega_0-{2a_1(t_0)\over H_0}+{\dot a_1(t_0)\over H_0^2}
={1\over 2}\Omega_0-{a_1(t_0)\over H_0}\left(2+{\Omega_0\over 2}\right)
-{k_1\over 2H_0^2}~.
\ee
This is no longer constrained to be positive. 

In Appendix \ref{proof2}, we show directly from the solutions to the Einstein equations, Eqs.\ (\ref{Rflat}), (\ref{para1}), and (\ref{para2}), that we must have both $t'_0(0)=0$ and $k'(0)=0$ in order to have a non-singular model, and if this is true then $q_0$ cannot be negative. In Appendix \ref{INNmodels} we use this to show that the models of Iguchi {\it et al.} have weak central singularities.

\subsection{Achieving a Negative Apparent Deceleration Parameter at Nonzero Redshifts}

Although models that have been previously analyzed contain central singularities, it is still possible to construct LTB models without such a singularity for which the effective deceleration parameter $q(z)$, as defined in Eq.\ (\ref{decel}), is negative for some nonzero redshifts. Here we explore a class of zero energy LTB models with a bang time function $t_{0}(r)$ that is quadratic near $r=0$, and therefore non-singular there.  

In a zero energy LTB model, we have
\be
dt=-R'(r,t)dr
\label{time}
\ee
and therefore we can get the equation for $t(r)$ along light rays that we observe from supernovae.  Also, $z$ is a function of $r$ via Eq.\ (\ref{redshift}), specialized to $k(r)=0$, and we get $z$ as a function of $r$ only by using our solution for $t(r)$ along the rays. The bang time function is chosen such that it will (i) approach a constant for large $r$, so as to have a uniform density for large redshifts, and (ii) have no terms linear in $r$, so as to avoid a singularity at the center.  Thus we integrate Eqs.\ (\ref{redshift}) and (\ref{time}) with the bang time function choice
\be
t_{0}(r)=-\frac{\lambda r_c r^2}{r^2+r_c^2 D^{2}}
\label{bangtime}
\ee
where $\lambda$ and $D$ are dimensionless parameters, $r_c=\left[6\pi G\rho_0(t_o)\right]^{-1/2}$, and we choose units where $r_c=1$. We choose the initial conditions at the center, $t(r=0)=1$ and $z(r=0)=0$, and we integrate from the center outward.  

\begin{figure}[h]
\includegraphics[width=\picwidth,height =8cm ]{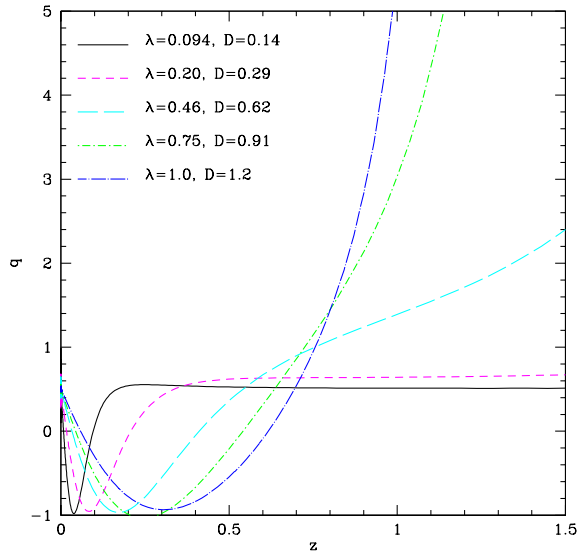}
\caption{The effective deceleration parameter $q$ versus redshift $z$ for several quadratic bang time models (\ref{bangtime}) which have a minimum $q$ of approximately negative one. Plotted here are the data for models with $(\lambda, D) = (0.094, 0.14),~(0.20, 0.29),~(0.46, 0.62),~(0.75, 0.91)$, and $(1.0, 1.2)$.} 
\label{quad} 
\end{figure}

Figure \ref{quad} displays results for the effective $q(z)$ that we calculate from the above model using Eqs.\ (\ref{lum}), (\ref{hubble}), and (\ref{decel}) for various values of $\lambda$ and $D$, namely $(\lambda, D) = (0.094, 0.14),~(0.20, 0.29),~(0.46, 0.62),~(0.75, 0.91)$, and $(1.0, 1.2)$.  We choose values of $\lambda$ and $D$ for which the minimum value of $q(z)$ that is attained is approximately $-1$.  As we can see, although all the models are forced to have $q(z=0)=1/2$, it is nevertheless possible for the deceleration parameter to become negative at nonzero redshifts, as we find a region of $q(z)<0$ for $z\alt 1$.

\begin{figure}[h]
\includegraphics[width=\picwidth,height =8cm ]{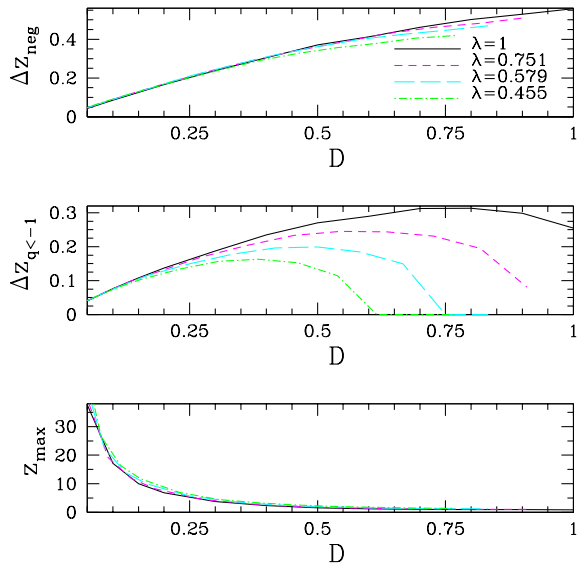}
\caption{Several measures of the feasibility of quadratic bang time models, plotted versus $D$ for $\lambda=1,~0.751,~0.589$, and $0.455$.  From top to bottom, we have plotted $\Delta z_{neg}$, $\Delta z_{q<-1}$, and $z_{max}$, all versus $D$.} 
\label{models}
\end{figure}

In order to reproduce the current luminosity distance data, we want $q(z)$ to quickly fall to from $q(0)=1/2$ to $q(z)\approx-1$ and then stay at that value until a redshift $z\sim 1$.  In Fig.\ \ref{models} we plot several quantities that encapsulate some of the characteristics of the functions $q(z)$, which are useful for assesing the feasibility of reproducing luminosity distance data.  We define $\Delta z_{neg}$ to be the width, in redshift, of the region where $q$ is negative, and $\Delta z_{q<-1}$ to be the width of the region where $q$ is below negative one.  We also found that the large redshift behavior is unstable in these models: $q$ blows up as we eventually approach the initial singularity. As an approximate measure of the location of this divergence, we define $z_{max}$ to be the redshift at which $q(z)$ exceeds 3.  Ideally, we want $\Delta z_{neg}\sim 1$, $\Delta z_{q<-1}=0$, and $z_{max}\rightarrow\infty$.  From Fig.\ \ref{models}, it does not appear as though this model can reproduce the data well, although it is conceivable that one could construct a model which gives more realistic results. We see that by increasing $D$, we also increase the size of the region with negative $q(z)$, which makes the model more phenomenologically viable; however, by increasing $D$, we also decrease $z_{max}$ and thus make the model less physically reasonable.

\section{The Inverse Problem}

Unfortunately, it is highly unlikely that one could guess a bang time function $t_{0}(r)$ that would yield the experimentally measured luminosity distance $D_{L}(z)$.  A better approach would be to solve the inverse problem: given the appropriate $D_{L}(z)$, work backwards to try to find the corresponding $t_{0}(r)$, which may or may not exist.  This approach has been taken before, but without avoiding the central singularity \cite{INN}, as is shown in Appendix \ref{INNmodels}. Models based on selected $D_{L}(z)$ generally break down at $z\sim 1$ upon encountering some pathology. We explore and clarify the possible pathologies below.

\subsection{General Properties}

In the LTB metric, the angular size distance is given by
\be
D_A(r,t)=R(r,t)=rT^2(r,t)~,
\ee
where $T\equiv[t-t_{0}(r)]^{1/3}$. Here we have specialized to units where $6\pi G\rho_0(t_o)=1$. We also define
the equivalent FRW radial coordinate to be
\be
\rfrw(z)\equiv\left(1+z\right)D_{A}(z)~,
\ee
in terms of which we have
\be
rT^2(1+z)=\rfrw(z)=\frac{D_{L}(z)}{1+z}~.
\label{frw}
\ee
Suppose we are given $\rfrw(z)$, and therefore $D_{L}(z)$ and $D_A(z)$, and from this we wish to find the corresponding zero energy LTB model.

The equations defining our flat LTB model with bang
time function may be written in the form
\be
\frac{dT}{dr}=-\frac{1}{3}+\frac{dt_{0}}{dr}\left(\frac{2r}{9T^{3}}-\frac{1}{3T^{2}}\right)
\label{alg1}
\ee
and
\be
\frac{1}{\left(1+z\right)}\frac{dz}{dr}=\left(\frac{2}{3T}+\frac{2r}{9T^{4}}\frac{dt_{0}}{dr}\right)~.
\label{alg2}
\ee
Multiply Eq.\ (\ref{alg1}) by $2/T$ and then add to Eq.\ (\ref{alg2}) to find
\be
\frac{dt_{0}}{dz}=\frac{3T}{2\left(1+z\right)\left(r/T-1\right)}\frac{d}{dz}\left[T^{2}\left(1+z\right)\right]~;
\ee
we can also combine Eqs.\ (\ref{alg1}) and (\ref{alg2}) such that we eliminate $dt_{0}/dr$ altogether to find
\be
\frac{r}{T}\frac{dT}{dz}+\left(\frac{r}{T}-1\right)\frac{dr}{dz}=\frac{1}{1+z}\left(r-\frac{3T}{2}\right)~.
\ee
Defining $X\equiv T^2(1+z)$, these equations can recast into
\be
{1\over X}{dX\over dz}
={(\rfrw\sqrt{1+z}/X^{3/2}-1)\over\rfrw(3/2-\rfrw\sqrt{1+z}/X^{3/2})}
\left[{3X^{3/2}\over 2(1+z)^{3/2}}-{d\rfrw\over dz}\right]
\label{Xeqn}
\ee
and 
\be
\dtzz={3X^{3/2}\over 2\rfrw(1+z)^{3/2}(3/2-\rfrw\sqrt{1+z}/X^{3/2})}
\left[{3X^{3/2}\over 2(1+z)^{3/2}}-{d\rfrw\over dz}\right]~.
\label{tzzeqn}
\ee
In the spatially flat, dust-dominated FRW model, $X=1$ and
$\rfrw(z)=3[1-1/\sqrt{1+z}]$.

Given $\rfrw(z)$, Eq.\ (\ref{Xeqn}) is a first order
ordinary differential equation for $X(z)$.
It becomes singular when
\be
{\rfrw(z)\sqrt{1+z}\over X^{3/2}}={3\over 2}~;
\label{critpt}
\ee
for a flat, dust-dominated FRW model, this occurs when $z=5/4$. Solutions $z=\zc$ of Eq.\ (\ref{critpt}), if these exist, are critical points of differential equation (\ref{Xeqn}).
Near the critical point,
\be
{1\over X}{dX\over dz}\approx{1\over 2(3/2-\rfrw\sqrt{1+z}/X^{3/2})}
\left[{1\over 1+z}-{d\ln\rfrw\over dz}\right]~
\ee
and
\be
\dtzz\approx {\rfrw(z)\over (1+z)(3/2-\rfrw\sqrt{1+z}/X^{3/2})}
\left[{1\over 1+z}-{d\ln\rfrw\over dz}\right]~.
\ee
Transcritical solutions, which are non-singular at the critical point, are possible provided that 
$(1+z)d\ln\rfrw/dz=1$ at the critical point. We discuss these solutions in
more detail below. Clearly, the spatially-flat, dust-dominated FRW model
is one special transcritical solution. For a general choice of $\rfrw(z)$, however, 
the conditions for passing smoothly through the critical point 
will not be generically satisfied, and both $d\ln X/dz$ and $dt_0/dz$ 
will diverge there. This suggests that a flat LTB model with
a bang time function can only mimic a generic $\rfrw(z)$ up to some
limiting redshift below $\zc$, where
\be
\kfrw(\zc)\equiv{1\over 1+\zc}-{d\ln\rfrw\over dz}\biggr\vert_{z=\zc}=0~.
\label{zcdef}
\ee
We shall argue below that only the special class of transcritical
solutions can extend to infinite redshift.

For exploring characteristics of the solutions, it proves
useful to define the new variable
\be
V\equiv1-{2\rfrw\sqrt{1+z}\over 3X^{3/2}}~.
\label{newvars}
\ee
Substituting Eq.\ (\ref{newvars}) into Eq.\ (\ref{Xeqn}) gives,
after some algebra,
\ba
{dV\over dz}&=&{(1-4V+V^2)\over 2V(1+z)}-{(1-V)^2\over 2V}{d\ln\rfrw
\over dz}\nonumber\\
&=&{(1+V^2)\over 2V}\left[{1\over 1+z}-{d\ln\rfrw\over dz}\right]
-{2\over 1+z}+{d\ln\rfrw\over dz}~.
\label{Veqn}
\ea
For a flat, dust-dominated FRW model, $V_{FRW}=3-2\sqrt{1+z}$, and substituting
this $V(z)$ into the right hand side of Eq.\ (\ref{Veqn}) yields $dV_{FRW}/dz=-1/\sqrt{1+z}$.
Near $z=0$, we have seen that flat LTB models resemble
flat, dust-dominated FRW models, so
\be
d\ln\rfrw/dz=z^{-1}[1-{1\over 2}(1+
q(0))z+\ldots]=z^{-1}(1-{3z/4}+\ldots)~,
\ee 
and therefore
\be
V(z)\approx 1-z+{z^2\over 4}-{z^3\over 8}+\ldots~
\label{vsmallz}
\ee
for $z\ll 1$. The first term in the small-$z$ expansion of $V(z)$ that
can deviate from Eq.\ (\ref{vsmallz}) is of order $z^4$.

At sufficiently small $z$, we expect $V(z)$ to decrease. There are three
possible classes of solutions to Eq.\ (\ref{Veqn}): (i) solutions that
decrease from $V(0)=1$ to some constant $\vinfty<1$ as $z\to\infty$, without
crossing the critical point at $V=0$; (ii) solutions that decrease
until a redshift $z=z_{0}<\zc$, where they terminate; and (iii) transcritical solutions
that pass through the critical point smoothly. We examine these three
classes in turn. In our considerations, we keep $\rfrw(z)$ general, with
the provisos that the model tends to $q=1/2$ at both $z\to 0$ and
$z\to\infty$. The former is dictated by the character of LTB
models free of central singularities, whereas the latter must be true
of any phenomenologically viable model. In particular, then, we assume
that $H\approx {2\over 3}\OF(1+z)^{3/2}$ at large $z$, where 
$\Omega_{\rm FRW}<1$. Therefore $\rfrw(z)\to\rfinfty$ as $z\to\infty$, where $\rfinfty$ is a constant.

Consider first solutions that decrease toward $V_\infty$ asymptotically.
At large values of $z$, Eq.\ (\ref{Veqn}) becomes
\ba
{dV\over dz}&\approx& {1-4\vinfty+\vinfty^2\over 2\vinfty(1+z)}
-{(1-\vinfty)^2\over 2\vinfty}{d\ln\rfrw\over dz}\nonumber\\
&\approx& {1-4\vinfty+\vinfty^2\over 2\vinfty(1+z)}
-{3(1-\vinfty)^2\over 4\vinfty\OF\rfinfty(1+z)^{3/2}}~.
\label{vassone}
\ea
The first term on the right hand side of Eq.\ (\ref{vassone}) is
negative as long as $\vinfty>V_0\equiv 2-\sqrt{3}\approx 0.27$,
and it dominates the second term. But $V(z)\sim -\ln(1+z)$ in that case,
and this diverges. Thus, we can only have $\vinfty=V_0$. In that
case, we let $V(z)=V_0+u(z)$ at large $z$, and find
\be
{du\over dz}\approx -{u\sqrt{3}\over(2-\sqrt{3})(1+z)}
-{3(\sqrt{3}-1)^2\over 4(2-\sqrt{3})\OF\rfinfty(1+z)^{3/2}}~,
\ee
which has the general solution
\be
u(z)={C\over (1+z)^{\sqrt{3}/(2-\sqrt{3})}}
-{3(\sqrt{3}-1)^2\over 2(3\sqrt{3}-2)\OF\rfinfty\sqrt{1+z}}~,
\ee
where $C$ is a constant. Although $u(z)\to 0$ as $z\to\infty$,
it approaches zero from below, not from above, which contradicts
our basic assumption. Thus, solutions that simply decrease
toward constant $V(z)>0$ asymptotically do not exist. Conceivably,
there can be solutions that decrease to a minimum and then
increase toward $V_0$ asymptotically. For these, however, $V(z)$
will be double valued. It then follows that $X=T^2(1+z)$ must
be double valued, since $\rfrw\sqrt{1+z}$ is monotonically 
increasing, and such behavior could be pathological. More generally,
we shall see below that solutions that avoid $V=0$ must 
terminate at $z=\zc$ in order to avoid other physical pathologies.
Thus, a solution ``on track'' to a minimum value $V_{\rm min}>0$, and on to $V_{0}$ asymptotically,
might end at finite redshift.

Next, consider solutions that reach $V=0$ at $z=z_0<\zc$ and
end there. Near $V=0$, Eq.\ (\ref{Veqn}) is approximately
\be
{dV\over dz}\approx{\kfrw(z_0)\over 2V}~,
\label{vnearzip}
\ee
where $\kfrw(z)=(1+z)^{-1}-d\ln\rfrw/dz$, as in Eq.\ (\ref{zcdef}).
Note that since $z_0<\zc$, $\kfrw(z_0)<0$ as well. The solution to Eq.\
(\ref{vnearzip}) is $V(z)\approx\sqrt{-\kfrw(z_0)(z_0-z)}$,
which terminates at $z=z_{0}$.

In order to reach
the critical point, we must have
\be
\kfrw(z)\leq 0
\ee
all the way up to the critical point, with equality holding at $z=\zc$ for 
the transcritical solution. For a transcritical solution to exist,
we must be able to expand
\be
\kfrw(z)= Q\Delta z+{\cal O}\left(\Delta z^2\right)
\label{Kexpand}
\ee
near the critical redshift, $\zc$, where $\Delta z=z-\zc$ and 
$Q>0$. For a flat, dust-filled FRW model, we have $\zc=5/4$ and $Q=8/27$. Using this linear approximation,
we find from Eq.\ (\ref{Veqn}) that $V=k\Delta z+{\cal O}(\Delta z^2)$, where the slope $k<0$ is the
solution to
\be
k^2+{k\over 1+\zc}-{Q\over 2}=0~.
\ee
That is, we need the negative root
\be
k=-{1\over 2(1+\zc)}-{1\over 2}\sqrt{\left({1\over 1+\zc}\right)^2
+2Q}~.
\label{ksoln}
\ee
For a flat, dust-filled FRW model, we find $k=-2/3$. This is clearly a transcritical
solution.

We can turn the above analysis into a test of whether a candidate
for $\rfrw(z)$ that agrees with observations can be represented
by a transcritical, zero energy LTB model.
First, for the candidate model, it is possible to
find $\zc$ and $Q$ algebraically; we can find $\zc$ using Eq.\ (\ref{zcdef}), by requiring
that $\kfrw(\zc)=0$,
and we can find $Q=dK_{FRW}/dz\vert_{z=\zc}$, cf.\ Eq.\ (\ref{Kexpand}). Next, find $k$
given $Q$ and $\zc$ from Eq.\ (\ref{ksoln}) and
use this value of $k$ to integrate Eq.\ (\ref{Veqn}) back toward $z=0$. If
the solution satisfies Eq.\ (\ref{vsmallz}) as $z\to 0$, 
then it is an acceptable
transcritical solution. 

There are other disasters that may befall the solution for 
general $\rfrw(z)$, and some of these may even afflict transcritical
solutions. Eq.\ (\ref{tzzeqn}) may be rewritten as
\be
\dtzz={2\rfrw\over 3(1+z)(1-V)}\left[{1\over V}\left({1\over 1+z}
-{d\ln\rfrw\over dz}\right)+{1\over (1+z)(1-V)}\right]~.
\label{newtzz}
\ee
As we have noted before, $dt_0/dz$ diverges at $V=0$ for generic $\rfrw(z)$,
but for transcritical solutions,
\be
\dtzz={2\rfrw(\zc)\over 3(1+\zc)}\left({Q\over k}+{1\over
1+\zc}\right)+{\cal O}\left(\Delta z\right)
\ee
near the critical point, which is finite, so this potential
disaster is avoided. In particular, for a flat,
dust-filled FRW model with $Q=8/27$ and $k=-2/3$ at $\zc=5/4$, we see
that $Q/k+{1/(1+\zc)}=0$, which is consistent with $t_0(z)=0$
for all redshifts.

We must check for two other possible disasters, for solutions
that are transcritical or not. As mentioned in the previous section, physical regions in any solution
must have a positive, finite $R'=\partial R/\partial r$ and $dr/dz$. We
find
\ba
{\partial R\over\partial r}&=&[3(1-V)]^{1/3}\left({2\rfrw\over 1+z}\right)^{2/3}
\left[{(1+z)d\ln\rfrw/dz-1\over
(1-V)[(1+z)d\ln\rfrw/dz-1]+2V}\right]\nonumber\\
{dr\over dz}&=&\left[{2\rfrw\over 3(1-V)(1+z)^4}\right]^{1/3}\left\{
\left({1-V\over 2V}\right)\left[(1+z){d\ln\rfrw\over dz}-1\right]
+1\right\}~.
\label{disasters}
\ea
Note that we have $\dot R'\propto (dr/dz)^{-1}$, and thus a finite, positive $dr/dz$
implies that $\dot R'>0$.
These equations are evaluated
along the path of a light ray directed radially inward. The
requirement $\partial R/\partial r>0$ along the light ray
is a necessary but not sufficient condition for an acceptable
model. The more general requirement is that $\partial R/\partial r>0$
at all $(r,t)$, a global condition that is much harder to satisfy; but in general, from Eq.\ (\ref{Rflat}), this will be satisfied
for models with $t_0(r)$ decreasing monotonically.
From the first of Eqs.\ (\ref{disasters}), we note that
$\partial R/\partial r\to 0$
at $z=\zc$ for solutions that are not transcritical. Solutions
that terminate at $z_0<\zc$ would not encounter this pathology.
Solutions of the first type described above, which decrease from
$V(0)=1$ but do not cross $V=0$, would end at $z=\zc$. For a transcritical
solution,
\be
{\partial R\over\partial r}=3^{1/3}\left({2\rfrw(\zc)\over 1+\zc}\right)^{2/3}
\left[{Q(1+\zc)\over Q(1+\zc)-2k}\right]>0
\ee
at the critical point. Transcritical solutions therefore propagate
right through the critical point with a positive, finite $\partial R/\partial
r$. From the second of Eqs.\ (\ref{disasters}), we see that 
$dr/dz$ diverges for solutions that terminate at $V=0$ and $z=z_0$.
For transcritical solutions
\be
{dr\over dz}=\left[{\rfrw(\zc)\over 3(1+\zc)}\right]^{1/3}
\left(-{Q\over k}+{2\over 1+\zc}\right)>0~
\ee
at the critical point.
Beyond the critical point, transcritical solutions have $V<0$, 
and for reasonable $\rfrw(z)$
with decreasing $(1+z)d\ln\rfrw/dz$,
it seems likely that $dr/dz$ remains positive. 

From these general considerations, we conclude that zero energy
LTB models can only mimic a given,
generic $\rfrw(z)$ -- arranged, for example, to fit observations
of Type Ia supernovae -- for $0\leq z\leq z_0<\zc$, where $\zc$ is
the solution to Eq.\ (\ref{zcdef}). There can be exceptional,
transcritical models that extend to infinite $z$ without any
mathematical or physical pathologies. However, transcritical
models are highly constrained mathematically, and may not
exist for choices of $\rfrw(z)$ that conform to phenomenological requirements.
The flat, dust-dominated FRW model is one transcritical solution, but it
is ruled out by observations.

\subsection{Manufacturing Transcritical Solutions}

To manufacture transcritical solutions, we will specify 
$V(\rtil)$, where $\rtil(z)\equiv\rfrw(z)\sqrt{1+z}$, and find
an equation for $\rtil(z)$. From Eq.\ (\ref{Veqn}) we find
\be
{d\rtil\over dz}={3-10V+3V^2\over 2(1+z)[2VV'+(1-V)^2/\rtil]}
\label{rtileqn}
\ee
where $V'(\rtil)\equiv dV(\rtil)/d\rtil$. As long as $VV'\to 0$
near $V=0$, Eq.\ (\ref{rtileqn}) satisfies the transcriticality
condition $\kfrw=0$ when $V=0$. As an example, suppose we assume
that $V=1-k\rtil$. Then we find $\rtil=(2/k)(\sqrt{1+z}-1)$ and
we must choose $k=2/3$ in order to have the proper behavior at small $z$.
This solution is simply equivalent to the flat, dust-filled FRW solution.

Superficially, the prescription is simple: specify a $V(\rtil)$,
make sure that $VV'\to 0$ when $V=0$, and then find the corresponding
$\rtil(z)$ by integrating Eq.\ (\ref{rtileqn}).  However, we know
that acceptable solutions must have $d\rtil/dz\geq 0$ and finite; 
these conditions are not so easy to guarantee.

Let us assume $V(\rtil)=1-{2\over 3}\rtil f(\rtil)$; then Eq.\
(\ref{rtileqn}) becomes
\be
{d\rtil\over dz}={(\rtil f-1)(\rtil f+3)\over
2(1+z)[f(\rtil f-1)-(1-2\rtil f/3)\rtil f']}~.
\label{rfeqn}
\ee
The numerator of
Eq.\ (\ref{rfeqn}) is zero when $\rtil f(\rtil)=1$, or $V=1/3$.
If the denominator of Eq.\ (\ref{rtileqn}) is nonzero at this point,
then $d\rtil/dz$ goes to zero, and changes sign upon crossing it.
Thus, we also want the denominator to vanish for an acceptable
solution. In other words, $V=1/3$ must be a critical point of
Eq.\ (\ref{rtileqn}): we have only succeeded in hiding the critical
nature of the problem, rather than eliminating it. Eq.\ (\ref{rfeqn})
shows that to pass through this critical point we must require
that $\rtil f'=0$ when $\rtil f=1$. Clearly, the spatially
flat, dust-filled FRW model, for which $f(\rtil)=1$, is one possibility.

It is also possible that the denominator of Eq.\ (\ref{rfeqn}) vanishes,
so $d\rtil/dz\to\infty$ before $\rtil f\to 1$. This happens when
\be
f'={f(\rtil f-1)\over \rtil (1-2\rtil f/3)}~.
\ee
If $f'<0$ at small values of $\rtil$, it is possible that infinite
$d\rtil/dz$ occurs before $\rtil f\to 1$. Since we also want 
$f(0)=1$ for nonsingular models, and $f(\infty)=$ constant for
models that approximate a flat FRW model with $\Omega_M<1$ at sufficiently
large redshift, we have several requirements
on $f(\rtil)$ that must be met simultaneously for a model that
is acceptable mathematically. Moreover, physically acceptable models
must also have $\Rp>0$ and $\Rdotp>0$. Only a subset of such models
-- if any -- will also be acceptable phenomenologically.

To examine the phenomenological properties of a candidate
transcritical solution, first define the effective Hubble
parameter $\heff(z)$ via
\be
{d\rfrw\over dz}={1\over\heff(z)}~.
\ee
It is straightforward to show that, normalizing so that $\heff(0)=1$,
\be
h(z)\equiv{\heff(z)\over (1+z)^{3/2}}={3\over 2}
\left[(1+z){d\rtil\over dz}-{\rtil\over 2}
\right]^{-1}={3[f(\rtil f-1)-(1-2\rtil f/3)\rtil f']\over 3(\rtil f-1)
+(1-2\rtil f/3)\rtil^2f'}~.
\label{heffdef}
\ee
We can also calculate the effective value of the equation of
state paraemter
\ba
\weff&=&{2(1+z)\over 3}{d\ln h\over dz}\nonumber\\
     &=&{(\rtil f+3)(\rtil f-1)\over 3[f(\rtil f-1)
          -(1-{2\over 3}\rtil f)\rtil f']^2
          [3(\rtil f-1)+(1-{2\over 3}\rtil f)\rtil^2 f']}
\nonumber\\
     & &\times\Biggl\{f'\left[{4(\rtil f)^3\over 3}+2(\rtil f)^2
              -8\rtil f+6\right]
     +(\rtil f')^2\left[-{2(\rtil f)^2\over 9}+{2\rtil f\over 3}+1
        \right]\nonumber\\& &
     -\rtil f''(\rtil f-1)(\rtil f+3)\left(1-{2\rtil f\over 3}\right)
     \Biggr\}~.
\label{weffdef}
\ea
$\heff(z)$ is the Hubble parameter that would be measured by
an observer who assumes her observations are described
by a spatially flat FRW model, and $\weff(z)$ is the associated
equation of state parameter. (A less dogmatic observer would
allow for the possibility of spatial curvature.) Note that Eq.\ (\ref
{heffdef}) implies that for $f'=0$, $\heff/(1+z)^{3/2}=f={\rm
constant}$. If we are interested in using LTB models
to mimic a spatially flat FRW model with a mixture of dust plus cosmological
constant, we shall want $f\to \sqrt{\Omega_M}=\sqrt{1-\Omega_V}$ 
at large redshift, where $\Omega_M$ and $\Omega_V$ are the present
density parameters in nonrelativistic matter and cosmological 
constant, respectively. Moreover, $f\to 1$ as $z\to 0$; thus 
$f$ must decrease from 1 to $\sqrt{1-\Omega_V}$ as redshift
increases to mimic observations in a FRW model of this sort. 

For any choice of $f(\rtil)$ tailored to pass smoothly
through both $V=0$ and $\rtil f(\rtil)=1$, we can
integrate Eq.\ (\ref{rfeqn}) to find a transcritical solution.
However, such a solution still must pass the tests outlined
in the previous section to extend to arbitrarily large redshifts.
In terms of $\rtil$ and $f(\rtil)$, 
\ba
{dt_0\over dz}&=&{3(h-f)\over 2(1+z)^{5/2}f^2h\rtil(1-2\rtil f/3)}\nonumber\\
(1+z){\partial R\over\partial r}&=&{f^{1/3}(3-2h\rtil)\over f+2h
-2fh\rtil}\nonumber\\
{3\dot R'\over 2(1+z)^{1/2}}&=&{3f^{1/3}(1-2\rtil f/3)\over 
2+f/h-2\rtil f}~.
\label{checkeqn}
\ea
We reject any transcritical solution for which $dt_0/dz$ is ever
positive, or $\partial R/\partial r$ or $\dot R'$ change sign. 
\footnote{
For LTB with bang time perturbations only, $(\partial R/\partial r)_t=
[t-t_0(r)]^{-1/3}[t-t_0(r)-{2\over 3}rdt_0/dr]$, which is only zero for
a shell at coordinate radius $r$
when $t-t_0(r)={2\over 3}r dt_0(r)/dr$. If $dt_0(r)/dr<0$, this occurs
before $t_0(r)$ and is therefore irrelevant. As long as $dt_0(z)/dz<0$ and
$\dot R'>0$ along the light ray path, $dt_0(r)/dr<0$ and $(\partial R/
\partial r)_t$ is never zero.}
Moreover, even if a transcritical solution is found that possesses
none of the pathologies discussed above, it may not conform to
observational constraints. Thus, if there are any 
nonsingular, non-pathological LTB solutions that can also mimic
the observations, they must be very exceptional indeed.

Designing nonsingular, nonpathological transcritical solutions
is a formidable challenge. Suppose that at small values of
$\rtil$, we expand $f(\rtil)=1+f_n\rtil^n+\ldots$. Then we find
that $h\approx 1+f_n(1+n)\rtil^n+\ldots$ and $\weff=f_nn(n+1)
\rtil^{n-1}+\ldots$ near the origin, where $\rtil\approx
{3\over 2}z$. Also 
\ba
\frac{dt_0}{dz} &\approx& \frac{3\left(h-f\right)}{2\rtil}\approx \frac{3}{2}f_nn\rtil^{n-1}\nonumber\\
\left(1+z\right)\frac{\partial R}{\partial r} &\approx& 1 - \frac{2\left(h-1\right)}{3} \nonumber\\
\frac{3\dot R'}{2(1+z)^{1/2}} &\approx& 1+\frac{1}{3}\left(h-1\right)~.
\ea
We wish to tailor $f(\rtil)$ to maintain positive values of both
$\Rp$ and $\Rdotp$, but already near the origin $\Rp$ and $\Rdotp$
deviate from their flat, dust-filled FRW relationships in opposite
senses. Notice that to avoid the weak singularity near
the origin, we need to have $n\geq 2$. Moreover, we want to make sure
that $t_0(z)$ is monotonically decreasing to avoid
shell crossing. Near the origin, decreasing $t_0(z)$ implies $f_n<0$.

To illustrate how difficult it is to manufacture non-pathological
transcritical solutions from Eq.\ (\ref{rfeqn}), we have considered 
\be
f(\rtil)={1\over\rtil_1}\left[1+K\left({\rtil_1^\pp-\rtil^\pp
\over \rtil_2^\pp+\rtil^\pp}\right)^p\right]~.
\label{fchoice}
\ee
The model has four parameters: $K$, $\rtil_1$, $\pp$, and $p$; the remaining parameter $\rtil_2$ will be determined in terms of these four.
To be sure that $h$ and $\weff$ are finite near $\rtil f=1$,
we want either $p\equiv 2$ or $p\geq 3$. 
Expanding near the origin, we find
\be
f(\rtil)={1+K(\rtil_1/\rtil_2)^\pp\over\rtil_1}
-{Kp\over\rtil_1}\left({\rtil_1\over \rtil_2}\right)^{\pp p}
\left({1\over \rtil_2^\pp}+{1\over\rtil_1^\pp}\right)\rtil^\pp
+\ldots~;
\ee
requiring $f(0)=1$ inplies
\be
\rtil_2^\pp=\rtil_1^\pp\left({K\over\rtil_1-1}\right)^{1/p}~,
\ee
and so we can rewrite the expansion in the form $f(\rtil)=1+f_n\rtil^n+\ldots$
with
\be
f_n=-{Kp\over\rtil_1}\left({\rtil_1\over \rtil_2}\right)^{\pp p}
\left({1\over \rtil_2^\pp}+{1\over\rtil_1^\pp}\right)
=-{p(\rtil_1-1)\over\rtil_1^{\pp+1}}\left[1+\left({\rtil_1-1\over K}
\right)^{1/p}\right]~.
\ee
Thus, we want $\rtil_1-1>0$ and therefore $K>0$ for $f_n<0$ and real $\rtil_2$. 
At large values of $\rtil$, $f(\rtil)\to \rtil_1^{-1}[1+(-1)^pK]$~. Thus,
we expect models that can mimic decelerating FRW models successfully
to have $\rtil_1^{-1}[1+(-1)^pK]<1$, suggesting either large $\rtil_1$
or odd $p$, or both.  Empirically, we have been unable to find
any non-pathological models based on Eq.\ (\ref{fchoice}) with these
properties.

\begin{figure}[h]
\centering
\includegraphics[width=\picwidth,height =8cm ]{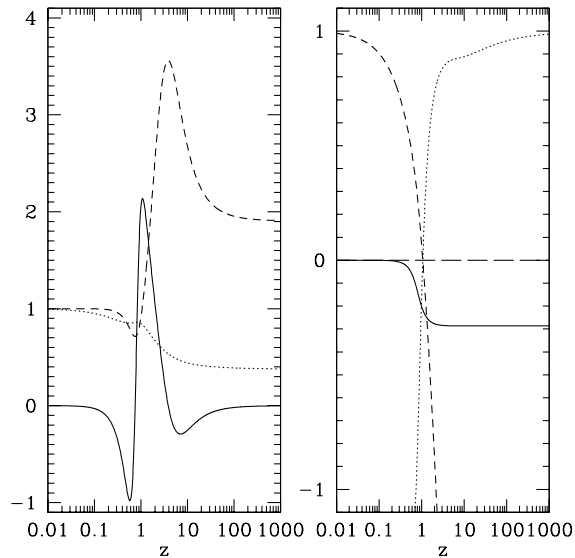}
\caption{Results for a transcritical model based on Eq.\ (\ref{fchoice})
with $K=1$, $p=2$, $\pp=3$, and $\rtil_1=1.05$, out to $z=1000$.
The left panel shows
$\rfrw(z)/\rfrw^{(0)}(z)$ (dotted), 
$h(z)\equiv H(z)/[H_0(1+z)^{3/2}]$ (short dashed), and $\weff(z)$ (solid). The reference model
corresponding to $\rfrw^{(0)}(z)$ is the spatially flat 
$\Lambda$CDM model with $\Omega_M=0.27$.
The right panel shows $V$ (short dashed), $1-d\ln\rfrw/d\ln(1+z)$ (dotted),
and $t_0$ (solid).}
\label{figtranspub}
\end{figure}

Figure \ref{figtranspub} shows an example of a transcritical solution
with $K=1$, $\rtil_1=1.05$, $\pp=3$, and $p=2$. Although the figure
only displays $z<1000$, we have integrated this model out to
$z=10^6$ to verify that it asymptotes smoothly to a high redshift
FRW model, with constant $t_0$. The left panel shows $\rfrw(z)/\rfrw^{(0)}(z)$ (dotted line), where $\rfrw^{(0)}(z)$
is computed for a flat $\Lambda$CDM FRW reference cosmology with 
$\Omega_M=0.27$ and $\Omega_\Lambda=0.73$, $h(z)$ (short
dashed line), and $\weff(z)$ (solid line).
Since $h\to 2/1.05=1.905$ at high $z$ for this
model, inevitably there must be regions with $\weff>0$; this is in
the range $0.75\lesssim z\lesssim 3.9$, with a peak value $\weff
\approx 2.13$. There are two regions of negative $\weff$: (i) one
at $0<z\lesssim 0.75$, with minimum value $\weff\approx -0.98$;
and (ii) an extensive regin at $z\gtrsim 3.9$, with a minimum
value $\weff\approx -0.292$. The right panel verifies the transcritical
nature of the solution: it shows $V$ (short dashed line), $1-d\ln\rfrw/
d\ln(1+z)$ (dotted line), and $t_0$ (solid line), and the long dashed line is
at $0$. The first two cross zero
simultaneously, as they must for a transcritical solution, and at
large redshifts, $V$ is approximately proportional to $\sqrt{1+z}$
while $1-d\ln\rfrw/d\ln(1+z)$ tends toward one. For this model,
$t_0(z)\leq 0$ at all $z$, and we have also verified that it
decreases monotonically. In addition, we can verify that the 
model behaves as predicted at small redshifts: $t_0(z)\approx
-0.34z^3$ and $\weff\approx -2.7z^2$.

\begin{figure}[h]
\centering
\includegraphics[width=\picwidth,height =8cm ]{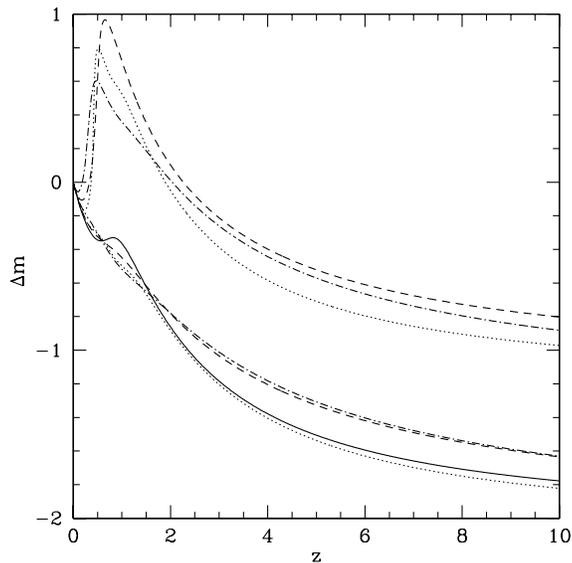}
\caption{Distance moduli relative to spatially flat
$\Lambda$CDM with $\Omega_M=0.27$ for models with $K=1$ and 
$(\pp, p)=(3, 2)$
(solid),  $(\pp, p)=(3, 4)$ (dotted),  $(\pp, p)=(2, 2)$
(dashed), and $(\pp, p)=(2, 4)$ (dash-dot), with $\rtil_1=
1.05$ for the lower set of curves, and $\rtil_1=1.5$ for
the upper set.
}
\label{transcomp}
\end{figure}

Figure \ref{transcomp} compares the relative distance moduli
\be
\Delta m=5.0\log_{10}[\rfrw(z)/\rfrw^{(0)}(z)]
\ee
for models with $K=1$ and $(\pp, p)=(3, 2)$
(solid line),  $(\pp, p)=(3, 4)$ (dotted line),  $(\pp, p)=(2, 2)$
(dashed line), and $(\pp, p)=(2, 4)$ (dash-dot line), with $\rtil_1
=1.05$ in the lower set
of curves and $\rtil_1=1.5$ in the upper set; there is
no solid line in the upper set for $(\pp, p)=(3, 2)$ because
the model is pathological.
For the lower set, luminous objects would appear systematically
brighter than they would in the standard $\Lambda$CDM model. As $\rtil_1$ is
increased, a period of substantial acceleration is seen in
the models below $z\sim 1$, leading to the systematic
brightening relative to $\Lambda$CDM as seen in the upper set
of curves. In either case, the luminosity differences would
be easy to discern observationally.

These few models illustrate several 
important qualitative points. First, it is possible to construct
non-pathological transcritical solutions that also avoid any
central weak singularities. Second, the model has a complicated
``effective equation of state'', including regions where $\weff<0$,
but also regions where $\weff>0$. In this case, we found a range of values
$-1\lesssim\weff\lesssim 2$. Finally, although it is possible
to construct models that are 
well-behaved mathematically, these models do not generally
conform to observational constraints. We have not, however, excluded the possibility that transcritical models in agreement with observations may exist.

\section{Conclusions}

Observations of Type Ia supernovae imply that we live in an accelerating universe, if interpreted within the framework of a homogeneous and isotropic cosmological model \cite{Riess, Perlmutter, Bennett}.  Some have tried to use the spherically symmetric LTB cosmological models to explain this seemingly anomalous data, as introducing a large degree of inhomogeneity can significantly distort the dependence of luminosity distance on redshift.  We have shown that one must take care in using these models, as they will contain a weak singularity at the symmetry center unless certain very restrictive conditions are met. Realistic LTB solutions require that the first derivative of the bang time function vanish at the center, $t_{0}'(0)=0$, and also that $k'(0)=0$, where $2E(r)\equiv-k(r)r^{2}$. Otherwise there are physical parameters, such as the density and Ricci scalar, which are not differentiable at the origin.

We have also shown that any LTB models without a central singularity will necessarily have a positive central deceleration parameter $q_{0}$, and thus all previously considered LTB models with $q_{0}<0$ are singular at the origin.  However, it is still possible to obtain a negative effective deceleration parameter for nonzero redshifts, which we have shown using as an example the model with zero energy and with the bang time function (\ref{bangtime}), that is quadratic at small $r$. These models have regions of apparent acceleration, where $q(z)$ is negative. If our goal is to reproduce luminosity distance data with a non-singular LTB model, we can try to smooth out the center appropriately and tailor the model to fit the data at modest redshifts, say $z\geq 0.01$. This is not an easy task because there are other singular behaviors that generally occur when trying to represent a given luminosity distance function $D_{L}(z)$ with a zero energy LTB model.

Our detailed examination of the ``inverse problem" elucidates how difficult it is to match zero energy LTB models to observed luminosity distance data.  We have shown that the underlying differential equations generically become singular at a critical point.  We have also shown that some exceptional choices of $\rfrw=D_{L}(z)/(1+z)$ admit transcritical solutions which are smooth at the critical point $z=\zc$, and may extend to arbitrarily high redshift, given that they do not encounter other pathologies along the way.  All other solutions terminate at some redshift $z_{0}<\zc$. We have shown how transcritical solutions can be constructed via a simple procedure. Although these solutions show both enhanced deceleration, seen as regions with $\weff(z)>0$, and acceleration, seen as regions with $\weff(z)<0$, none that we have constructed explicitly conform to observations. Here we have only studied the effects of a bang time function, and did not consider the case of a non-zero energy function $E(r)$.  We expect generic solutions with $E(r)\neq 0$ to share the basic characteristics of the models studied here, namely the critical points and other singularities that we have discussed.  However, we cannot say for sure that there are no transcritical and nonsingular solutions with non-zero $t_{0}(r)$ and $E(r)$ that agree with observational data on $r_{FRW}(z)$, although it does not appear to be likely, as is evident from previous unsuccessful attempts to find such solutions \cite{INN}.

Even if it were possible to reproduce determinations of $D_{L}(z)$ from supernova data in a LTB model without dark energy, we would still be left with the task of matching all of the other cosmological data with such a model.  First, the Wilkinson Microwave Anisotropy Probe is one of our most important sources of information about the Universe, via CMBR data; Alnes {\it et al.} \cite{Alnes2} try to reproduce the first peak of the angular power spectrum with LTB models, and Schneider and C\'{e}l\'{e}rier \cite{Celerier2} claim to be able to account for the apparent anisotropy in the dipole and quadrupole moments with an off center observer.  There are further constraints on inhomogeneous models from the kinetic Sunyaev-Zel'dovich effect, which constrains radial velocities relative to the CMB \cite{Benson}.  However, observations of large scale structure formation may be the most difficult to reconcile.  These data strongly disfavor a currently dust dominated universe, as density perturbations would have grown too much without dark energy present to speed up the cosmic expansion rate and consequently retard the growth of fluctuations.

\begin{acknowledgments}
This research was supported in part by NSF grants AST-0307273 and PHY-0457200.
\end{acknowledgments}

\appendix

\section{Proof of $q_0\geq 0$ directly from LTB solutions}
\label{proof2}

In this appendix we show directly from the solutions (\ref{Rflat}), (\ref{para1}), and (\ref{para2}) of the Einstein equations that LTB models without central singularities must have positive $q_{0}$. The zero energy solution $k(r)=0$ has
\be
a_1(t)\propto R''(r=0,t)=-\frac{4}{3}\left[6\pi G\rho_{0}\left(t_{o}\right)\right]^{1/3}t_{0}'(0)t^{-1/3}~.
\label{condition}
\ee
Thus, we see that the zero energy solution requires $t_{0}'(0)=0$ in order to have no central singularity.  More generally, for the $k(r)>0$ solutions, we find
\be
R''(0,t)=k'(0)\left[\frac{3F'\left(x_{0}\right)}{\sqrt{k_{0}}}t-\frac{8\pi G\rho_{0}\left(t_{o}\right)F\left(x_{0}\right)}{3k_{0}^{2}}\right]-t_{0}'(0)\left[2\sqrt{k_{0}}F'\left(x_{0}\right)\right]
\label{pos}
\ee
where $k_{0}\equiv k(r=0)$ and we have defined the function $F(x)$ by
\be
1-\cos u\equiv F(u-\sin u)~.
\ee
Here $x_{0}(t)=u_{0}-\sin u_{0}$ is the value of $x$ at the center $r=0$ at time $t$:
\be
x_{0}(t)=\frac{3k_{0}^{3/2}}{4\pi G\rho_{0}\left(t_{o}\right)}t~.
\ee
Similarly, for the $k(r)<0$ solutions we find
\be
R''(0,t)=-k'(0)\left[\frac{3G'\left(x_{0}\right)}{\sqrt{-k_{0}}}t-\frac{8\pi G\rho_{0}\left(t_{o}\right)G\left(x_{0}\right)}{3k_{0}^{2}}\right]-t_{0}'(0)\left[2\sqrt{-k_{0}}G'\left(x_{0}\right)\right]
\label{neg}
\ee
where we define the function $G(x)$ by
\be
\cosh u-1\equiv G(\sinh u-u)~.
\ee
Since the bracketed expressions in Eqs.\ (\ref{pos}) and (\ref{neg}) are functions of time, $R''(r=0,t)$ will vanish at arbitrary $t$ only if $t_{0}'(0)=0$ and $k'(0)=0$, and only then can one avoid having a singularity at the symmetry center.

These conditions, $t_{0}'(0)=0$ and $k'(0)=0$, lead to the restriction that $q_{0}$ must be positive.  C\'{e}l\'{e}rier \cite{Celerier} expands the luminosity distance for small redshift and finds the second order coefficient to be
\be
D_{L}^{(2)}\equiv\frac{1}{2}\left[\frac{d^{2}D_{L}}{dz^{2}}\right]_{r=0}=\frac{1}{2}\left[\frac{R'}{\dot{R}'}\left(1+\frac{R'\ddot{R}'}{\dot{R}'^{2}}+\frac{R''}{R'\dot{R}'}-\frac{\dot{R}''}{\dot{R}'^{2}}\right)\right]_{r=0}~,
\ee
where overdots again denote partial derivatives with respect to time. The deceleration parameter at $r=0$ is therefore
\be
q_{0}=1-2H_{0}D_{L}^{(2)}=\left[-\frac{R'\ddot{R}'}{\dot{R}'^{2}}-\frac{R''}{R'\dot{R}'}+\frac{\dot{R}''}{\dot{R}'^{2}}\right]_{r=0}~.
\ee
If $R''(0,t)=0$ to avoid a singularity, we find that the last two terms in the above expression are also zero, and we obtain
\be
q_{0}=\left[-\frac{R'\ddot{R}'}{\dot{R}'^{2}}\right]_{r=0}~.
\ee
Since $R'(r,t)>0$ to prevent shell crossing, and $\dot{R}^{2}$ is obviously positive, we would need to have 
\be
\ddot{R}'(0,t)=\ddot{a}_{0}(t)>0 
\ee
in order to have a negative $q_{0}$.  For the $k(r)=0$ solution,
\be
\ddot{R}'(0,t)=-\frac{2}{3}\left[\frac{2\pi G\rho_{0}\left(t_{o}\right)}{9}\right]^{1/3}t^{-4/3}<0 ~;
\ee
moreover, the $k(r)>0$ solution has
\be
\ddot{R}'(0,t)=-\frac{4\pi G\rho_{0}\left(t_{o}\right)}{3k_{0}}\left(\frac{du_{0}}{dt}\right)^{2}<0~,
\ee
and the $k(r)<0$ solution has
\be
\ddot{R}'(0,t)=\frac{4\pi G\rho_{0}\left(t_{o}\right)}{3k_{0}}\left(\frac{du_{0}}{dt}\right)^{2}<0~.
\ee
Therefore we can conclude that, in the absence of weak central singularities, all LTB solutions have positive $q_{0}$ since $\ddot{R}'(0,t)$ is always negative.

\section{Models of Iguchi, Nakamura and Nakao}
\label{INNmodels}

In this appendix we verify explicitly that the models with $q_0<0$ studied by Iguchi {\it et al.} \cite{INN} contain weak singularities. For the first case in Iguchi {\it et al.}, the pure bang time inhomogeneity, there will be no singularity if $t_{0}'(0)=0$, as shown from Eq.\ (\ref{condition}).  If we expand $D_{L}(z)$ for this FRW model in a power series around $z=0$, we can compare this term by term to the expansion of the luminosity distance for a zero energy LTB model to find \cite{Celerier}
\be
\Omega_{M}=1+5\frac{t_{0}'(0)}{\alpha\beta^2}+\frac{29}{4}\frac{t_{0}'^2(0)}{\alpha^2\beta^4}+\frac{5}{2}\frac{t_{0}''(0)}{\alpha^2\beta}
\label{omegam}
\ee
and
\be
\Omega_{\Lambda}=-\frac{1}{2}\frac{t_{0}'(0)}{\alpha\beta^2}+\frac{29}{8}\frac{t_{0}'^2(0)}{\alpha^2\beta^4}+\frac{5}{4}\frac{t_{0}''(0)}{\alpha^2\beta}~,
\label{omegal}
\ee
where $\alpha\equiv [6\pi G \rho_0(t_o)]^{1/3}$ and $\beta\equiv t^{1/3}$. Using the fact that $\Omega_{M}+\Omega_{\Lambda}=1$, we combine Eqs.\ (\ref{omegam}) and (\ref{omegal}) to find
\be
t_{0}'(0)=-\frac{1}{2}\alpha\beta^2\Omega_{\Lambda}~.
\ee
A nonzero $\Omega_{\Lambda}$ requires that $t_{0}'(0)$ is also nonzero, and hence there will be a singularity in such models.  

Iguchi {\it et al.} also look at models with $t_{0}(r)=0$ and positive $E(r)$.  By combining and rearranging Eqs.\ (6) and (39) from \cite{Celerier}, we find that
\be
\frac{3\Omega_{\Lambda}-1}{2}=\frac{R'\ddot{R}'}{\dot{R}'^{2}}+\frac{R''}{R'\dot{R}'}-\frac{\dot{R}''}{\dot{R}'^{2}}~.
\ee
Plugging into this the negative $k$ solution for $R(r,t)$ and then setting $r=0$, we can find an equation for $k'(0)$. Iguchi {\it et al.} make some simplifying definitions, wherein they set $H_{0}=G=1$ and then write everything else as a function of a parameter $\Omega_{0}$, which they vary between 0.1 and 1.  They set $k_{0}=\Omega_{0}-1$, $\rho_{0}(t_o)=3\Omega_{0}/8\pi$,
\be
u_{0}=\ln\left[\frac{2-\Omega_{0}}{\Omega_{0}}+\sqrt{\left(\frac{2-\Omega_{0}}{\Omega_{0}}\right)^{2}-1}\right]~,
\ee
and
\be
t(r=0)=\frac{\Omega_{0}}{2}\frac{\left(\sinh u_{0}-u_{0}\right)}{\left(1-\Omega_{0}\right)^{3/2}}~,
\ee
where $t(r)$ is evaluated along radially inward-moving light rays. Plugging these in and then solving for $k'(0)$ yields
\be
k'(0)=\frac{\left(1-\Omega_{0}\right)^{3/2}}{6}\left[\frac{\left(3\Omega_{\Lambda}-1\right)\sinh^{2}u_{0}\left(\cosh u_{0}-1\right)+2\left(\cosh u_{0}-1\right)^2}{3\sinh u_{0}-u_{0}\left(\cosh u_{0}+2\right)}\right]~,
\ee
where it is assumed that $\Omega_{\Lambda}=0.7$.  This shows that $k'(0)$ is only zero if $\Omega_{0}=1$; we can see from their Fig.\ (4) that this corresponds to the uninteresting FRW dust solution $k(r)=0$ for all $r$.  All of the other choices for $\Omega_{0}$ will correspond to models with $k'(0)\neq 0$ and a central singularity. Therefore, all of the non-trivial models computed in Ref.\ \cite{INN} have weak singularities at $r=0$.

\end{document}